\documentclass[twocolumn]{aastex63}
\usepackage{natbib}
\usepackage{amsmath}
\usepackage{bm}

\shortauthors{Saha \& Sengupta}
\shorttitle{Critical analysis of TESS data for five transiting exoplanets}

\begin{document}
	
	\title{CRITICAL ANALYSIS OF TESS TRANSIT PHOTOMETRIC DATA: IMPROVED PHYSICAL PROPERTIES FOR FIVE EXOPLANETS}
	
	\author[0000-0001-8018-0264]{Suman Saha}
	\affiliation{Indian Institute of Astrophysics, II Block, Koramangala, Bengaluru, India}
	\affiliation{Pondicherry University, R.V. Nagar, Kalapet, Puducherry, India}
	
	\correspondingauthor{Suman Saha}
	\email{suman.saha@iiap.res.in}
	
	\author[0000-0002-6176-3816]{Sujan Sengupta}
	\affiliation{Indian Institute of Astrophysics, II Block, Koramangala, Bengaluru, India}
	
	\accepted{for publication in The Astronomical Journal}
	
	\begin{abstract}
		
		We present improved physical parameters for four hot Jupiters: KELT-7 b, HAT-P-14 b, WASP-29 b, WASP-95 b, and a hot Neptune: WASP-156 b, by performing critical and rigorous analysis of the time-series observations from the Transiting Exoplanet Survey Satellite (TESS). Being a space-based telescope, the transit photometric data obtained by TESS are free from any noise component due to the interference of Earth's atmosphere. In our analysis of the observed data, we have used critical noise reduction techniques, e.g., the wavelet denoising and Gaussian process regression, in order to effectively reduce the noise components that arise from other sources, such as various instrumental effects and the stellar activity and pulsations. The better quality of photometric data from TESS, combined with our state-of-the-art noise reduction and analysis technique, has resulted into more accurate and precise values of the physical properties for the target exoplanets than that reported in earlier works.
		
	\end{abstract}
	
	
	
	\section{Introduction}
	
	The transit photometric follow-up studies of known exoplanets are aimed at improving their physical properties through repeated observations using better facilities and by using improved noise reduction and analysis techniques \citep{2011ApJ...726...94C, 2017AA...601A..53E, 2019AJ....157..102L, 2019AJ....158...39C, Saha_2021}. The results from the follow-up studies over a prolonged period of time can also be used to study the planetary dynamics and the presence of other undiscovered planetary mass objects in those systems \citep{nesvorny2012detection, 2015ApJ...810L..23J, gillon17, patra17, maciejewski18}. The improved values of various planetary parameters obtained through follow-up observations and precise noise reduction and data analysis would help interpreting the observed transmission spectra of the exoplanets more accurately through precise modeling \citep{2020ApJ...889..181S, 2020ApJ...898...89C}.
	
	\begin{deluxetable*}{LCCCCC}
		\tablecaption{Physical properties of the host stars\label{tab:tab1}}
		\tablewidth{0pt}
		\tablehead{ & \colhead{KELT-7 } & \colhead{HAT-P-14 } & \colhead{WASP-29 } & \colhead{WASP-95 } & \colhead{WASP-156 }}
		\startdata
		K_{RV}\;[m\:s^{-1}] & 138\pm19 & 219\pm3.3 & 35.6\pm2.7 & 175.7\pm1.7 & 19\pm1 \\
		R_\star\;[R_\sun] & 1.715\pm0.049 & 1.468\pm0.054 & 0.808\pm0.044 & 1.13\pm0.08 & 0.76\pm0.03 \\
		M_\star\;[M_\sun] & 1.483\pm0.069 & 1.386\pm0.045 & 0.825\pm0.033 & 1.11\pm0.09 & 0.842\pm0.052 \\
		T_{eff}\;[K] & 6789\pm49 & 6600\pm90 & 4800\pm150 & 5830\pm140 & 4910\pm61 \\
		\text{Sources} & \text{\citet{2015AJ....150...12B}} & \text{\citet{2010ApJ...715..458T}} & \text{\citet{2010ApJ...723L..60H}} & \text{\citet{2014MNRAS.440.1982H}} & \text{\citet{2018AA...610A..63D}}
		\enddata
	\end{deluxetable*}
		 	
	The transit follow-up observation from the space-based telescopes has an edge as they are unaffected by the variability of Earth's atmosphere. Another advantage of space-based telescopes is their longer available observational time as compared to their ground based counter-parts that remain dormant owing to daylight and unfavorable weather conditions. This longer available observation time is an useful aspect in continuous monitoring and follow-up of successive transit events.	The Transiting Exoplanet Survey Satellite (TESS) \citep{2015JATIS...1a4003R} is a space-based survey mission for the discovery of new exoplanets around bright nearby stars. It is also a powerful tool for transit follow-up studies of already discovered exoplanets, because it covers a large portion ($\sim75\%$) of the sky over the span of the entire survey. Space based instruments as TESS are affected by the scattered light from the Earth and the Moon, which causes increased background sky levels during certain parts of certain sectors. Also, though the observations from TESS is deprived of all noise components that arise by the interference of Earth's atmosphere, the noise components due to various instrumental effects and the stellar activity and pulsations need to be effectively reduced in order to estimate the physical properties of the target exoplanets with better accuracy and precision.
	
	We have recently reported \citep{Saha_2021} multiband transit follow-up studies of a few exoplanets with the implementation of critical noise reduction techniques that provided improved physical properties. The present study is a continuation of that project. Here we have used the photometric time-series data from TESS to conduct transit follow-up studies of four hot Jupiters: KELT-7 b \citep[e.g.,][]{2015AJ....150...12B}, HAT-P-14 b \citep[e.g.,][]{2010ApJ...715..458T}, WASP-29 b \citep[e.g.,][]{2010ApJ...723L..60H}, WASP-95 b \citep[e.g.,][]{2014MNRAS.440.1982H}, and a hot Neptune: WASP-156 b \citep[e.g.,][]{2018AA...610A..63D}. The instrumental effects of TESS contribute to the small scale fluctuations in the light-curves, which are uncorrelated in time. These fluctuations are similar in effect to the fluctuations in the light-curves from the ground-based telescopes due to the effect of Earth's atmosphere \citep{2019AJ....158...39C, Saha_2021}, but differ in magnitude scale, which is towards the lower end for TESS. We have used the wavelet denoising technique \citep{Donoho1994IdealDI, 806084, WaveletDenoise2012, 2019AJ....158...39C, Saha_2021} to effectively reduce this noise component without compromising the high-frequency components of the transit signals. The stellar activity and pulsations result in small temporal scale correlated noise in the light-curves. These noise components are treated by modeling them simultaneously along with the transit signal using the Gaussian Process (GP) regression method \citep{2006gpml.book.....R, 2015ApJ...810L..23J, 2019MNRAS.489.5764P, 2020AA...634A..75B, 2019AJ....158...39C, Saha_2021}. The improvements of the estimated physical properties due to the use of these critical noise reduction techniques is also demonstrated (see section \ref{sec:dis}).
	
	Following \citet{2019AJ....158...39C, Saha_2021}, we have modeled the transit light-curves by using the analytical transit formalism given by \citet{2002ApJ...580L.171M}. This incorporates the quadratic limb-darkening effect and uses the Markov chain Monte Carlo (MCMC) sampling technique by invoking the Metropolis-Hastings algorithm \citep{1970Bimka..57...97H}. From this modeling, we have estimated the transit parameters for our target exoplanets and used them to estimate the ephemeris parameters and other derivable parameter. For a streamlined flow of data analysis and modeling, we have used our semi-automated software package, previously used in \citet{Saha_2021}. We have compared the estimated parameter values obtained from this study with those from the previous studies (see section \ref{sec:dis}) to determine the improvement brought upon by this work.
	
	In section \ref{sec:obs}, we have detailed target selection and observations. In section \ref{sec:analysis}, we have described data analysis and modeling procedures. In section \ref{sec:dis}, we have discussed the significance of our results and in section \ref{sec:con}, we have derived conclusions.
	
	\begin{figure*}
		\centering
		\includegraphics[width=0.8\linewidth]{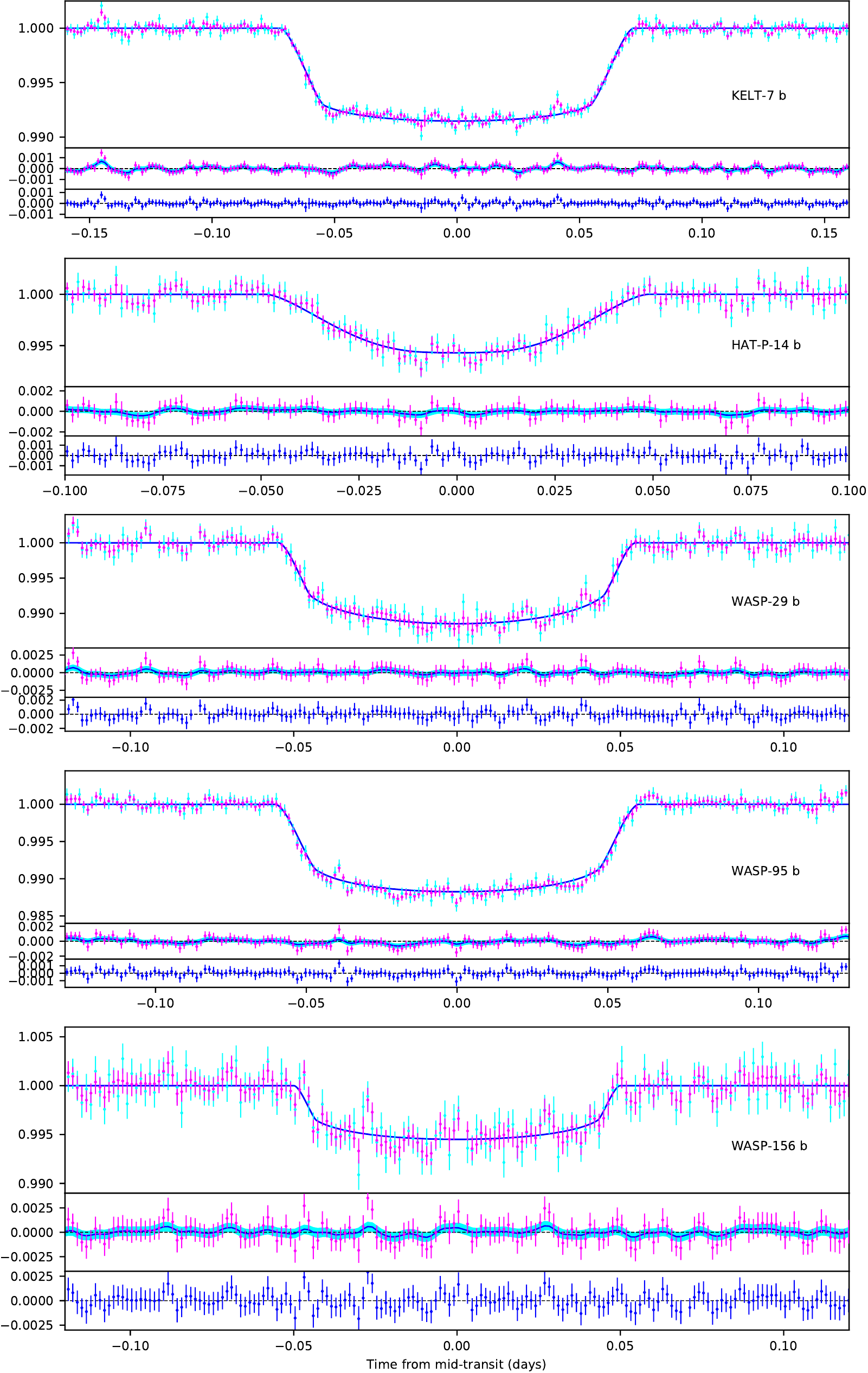}
		\caption{Observed and best-fit model light-curves of one transit event for each of our target exoplanets. For each observed transit, Top: the unprocessed light-curve (cyan), light-curve after wavelet denoising (magenta), the best-fit transit model (blue). Middle: the residual after modeling without GP regression (magenta), the mean (blue) and 1-$\sigma$ interval (cyan) of the best-fit GP regression model. Bottom: mean residual flux (blue). All lightcurves used in this study are shown in Figures \ref{fig:fig1}-\ref{fig:fig5}}
		\label{fig:fig0}
	\end{figure*}
	
	\begin{deluxetable*}{LCCC}
		\tablecaption{Estimated physical parameters for KELT-7 b \label{tab:tab2}}
		\tablewidth{0pt}
		\tablehead{\colhead{Parameter} & \colhead{This work} & \colhead{Exp values$^{\:\bm\dag}$} & \colhead{\citet{2015AJ....150...12B}}}
		\startdata
		\text{Transit parameters}\\
		T_0\; [BJD_{TDB}] & 2458816.518385\pm0.000051 & 2458816.518344\pm0.000083 & 2456352.495016\pm0.000191\\
		P\; [days] & 2.734784\pm0.000009 & 2.734788\pm0.000015 & 2.7347795\pm0.0000037\\
		b & 0.59178_{-0.00039}^{+0.00036} & 0.59185_{-0.00082}^{+0.00072} & 0.599_{-0.026}^{+0.023}\\
		R_\star/a & 0.178949_{-0.000020}^{+0.000022} & 0.178968_{-0.000093}^{+0.000078} & 0.1828\pm0.0040\\
		R_p/R_\star & 0.089597_{-0.000019}^{+0.000021} & 0.089623_{-0.000057}^{+0.000071} & 0.09074_{-0.00066}^{+0.00067}\\
		\text{Limb-darkening coefficients}\\
		C_1 & 0.2106_{-0.0049}^{+0.0035} & 0.2078_{-0.0075}^{+0.0068} & \nodata\\
		C_2 & 0.2192_{-0.0055}^{+0.0059} & 0.218_{-0.011}^{+0.010} & \nodata\\
		\text{Derived parameters}\\
		T_{14}\; [hr] & 3.4555_{-0.0010}^{+0.0012} & 3.4557_{-0.0023}^{+0.0025} & 3.511_{-0.022}^{+0.023}\\
		a/R_\star & 5.58818_{-0.00069}^{+0.00061} & 5.5876_{-0.0024}^{+0.0029} & 5.47\pm0.12\\
		i\; [deg] & 83.9211_{-0.0038}^{+0.0039} & 83.9203_{-0.0096}^{+0.0084} & 83.72_{-0.39}^{+0.40}\\
		M_p\; [M_J] & 1.24_{-0.17}^{+0.18} & 1.24_{-0.17}^{+0.18} & 1.28\pm0.18 \\
		M_p\; [M_\earth] & 394.2_{-55.1}^{+55.7} & 394.2_{-55.1}^{+55.7} & 407\pm57 \\
		T_{eq}\; [K] & 2030.8\pm14.6 & 2030.8\pm14.6 & 2051_{-27}^{+28}\\
		a\; [AU] & 0.0446\pm0.0013 & 0.0446\pm0.0013 & 0.04364_{-0.00068}^{+0.00067}\\
		R_p\; [R_J] & 1.293\pm0.060 & 1.293\pm0.060 & 1.514_{-0.050}^{+0.051}\\
		R_p\; [R_\earth] & 14.49\pm0.67 & 14.49\pm0.67 & 16.97_{-0.56}^{+0.57}\\
		\enddata
		\tablecomments{$^{\:\bm\dag}$ values estimated without using wavelet denoising and GP regression techniques (for comparison purpose).}
	\end{deluxetable*}
	
	\begin{deluxetable*}{LCCC}
		\tablecaption{Estimated physical parameters for HAT-P-14 b \label{tab:tab3}}
		\tablewidth{0pt}
		\tablehead{\colhead{Parameter} & \colhead{This work} & \colhead{Exp values$^{\:\bm\dag}$} & \colhead{\citet{2010ApJ...715..458T}}}
		\startdata
		\text{Transit parameters}\\
		T_0\; [BJD_{TDB}] & 2458984.653413\pm0.000047 & 2458984.653370\pm0.000093 & 2454875.28938\pm0.00047\\
		P\; [days] & 4.627686\pm0.0000077 & 4.627694\pm0.000015 & 4.627669\pm0.000005\\
		b & 0.91039_{-0.00019}^{+0.00017} & 0.91044_{-0.00033}^{+0.00034} & 0.891_{-0.008}^{+0.007}\\
		R_\star/a & 0.112595_{-0.000046}^{+0.000039} & 0.112590_{-0.000061}^{+0.000066} & 0.1127\pm0.0037\\
		R_p/R_\star & 0.081607_{-0.000024}^{+0.000020} & 0.081609_{-0.000063}^{+0.000067} & 0.0805\pm0.0015\\
		\text{Limb-darkening coefficients}\\
		C_1 & 0.2486_{-0.0072}^{+0.0080} & 0.150_{-0.011}^{+0.012} & 0.1089\\
		C_2 & 0.2805_{-0.0079}^{+0.0084} & 0.281_{-0.015}^{+0.013} & 0.2439\\
		\text{Derived parameters}\\
		T_{14}\; [hr] & 2.3386_{-0.0013}^{+0.0014} & 2.3384\pm0.0021 & 2.1888\pm0.0408\\
		a/R_\star & 8.8881_{-0.0031}^{+0.0036} & 8.8818_{-0.0052}^{+0.0048} & 8.87\pm0.29\\
		i\; [deg] & 84.1167_{-0.0025}^{+0.0026} & 84.1164_{-0.0046}^{+0.0050} & 83.5\pm0.3\\
		M_p\; [M_J] & 2.244\pm0.059 & 2.244\pm0.059 & 2.232\pm0.059\\
		M_p\; [M_\earth] & 713.1_{-18.7}^{+18.8} & 713.1_{-18.6}^{+18.9} & 709.4\pm18.7\\
		T_{eq}\; [K] & 1566.0\pm21.2 & 1566.0\pm21.2 & 1570\pm34\\
		a\; [AU] & 0.0606\pm0.0022 & 0.0606\pm0.0022 & 0.0606\pm0.0007\\
		R_p\; [R_J] & 1.101\pm0.036 & 1.101\pm0.036 & 1.150\pm0.052\\
		R_p\; [R_\earth] & 12.34\pm0.40 & 12.34\pm0.40 & 12.89\pm0.58\\
		\enddata
		\tablecomments{$^{\:\bm\dag}$ values estimated without using wavelet denoising and GP regression techniques (for comparison purpose).}
	\end{deluxetable*}
	
	\begin{deluxetable*}{LCCC}
		\tablecaption{Estimated physical parameters for WASP-29 b \label{tab:tab4}}
		\tablewidth{0pt}
		\tablehead{\colhead{Parameter} & \colhead{This work} & \colhead{Exp values$^{\:\bm\dag}$} & \colhead{\citet{2010ApJ...723L..60H}}}
		\startdata
		\text{Transit parameters}\\
		T_0\; [BJD_{TDB}] & 2458356.414870\pm0.000032 & 2458356.414808\pm0.000066 & 2455320.2341\pm0.004\\
		P\; [days] & 3.92271218\pm0.00000025 & 3.92271254\pm0.00000053 & 3.922727\pm0.000004\\
		b & 0.1160_{-0.0039}^{+0.0036} & 0.1169_{-0.0077}^{+0.0078} & 0.29\pm0.15\\
		R_\star/a & 0.080005_{-0.000040}^{+0.00044} & 0.080008_{-0.000061}^{+0.000074} & \nodata\\
		R_p/R_\star & 0.096602_{-0.000049}^{+0.000039} & 0.096595_{-0.000069}^{+0.000056} & 0.1010\pm0.0020\\
		\text{Limb-darkening coefficients}\\
		C_1 & 0.5001_{-0.0032}^{+0.0046} & 0.4995_{-0.0077}^{+0.0069} & \nodata\\
		C_2 & 0.1396_{-0.0037}^{+0.0038} & 0.1406_{-0.0076}^{+0.0087} & \nodata\\
		\text{Derived parameters}\\
		T_{14}\; [hr] & 2.6178_{-0.0014}^{+0.0016} & 3.0881_{-0.0034}^{+0.0039} & 2.6592\pm0.036\\
		a/R_\star & 12.4992_{-0.0069}^{+0.0063} & 12.499_{-0.012}^{+0.010} & \nodata\\
		i\; [deg] & 89.468{-0.017}^{+0.018} & 89.464_{-0.036}^{+0.035} & 88.8\pm0.7\\
		M_p\; [M_J] & 0.243_{-0.019}^{+0.020} & 0.257\pm0.020 & 0.244\pm0.020\\
		M_p\; [M_\earth] & 77.2_{-6.1}^{+6.2} & 81.5_{-6.5}^{+6.6} & 77.547\pm6.356\\
		T_{eq}\; [K] & 960.0_{-29.8}^{+29.9} & 960.1_{-29.9}^{+29.8} & \nodata\\
		a\; [AU] & 0.0470\pm0.0025 & 0.0470\pm0.0025 & 0.0457\pm0.0006\\
		R_p\; [R_J] & 0.775\pm0.031 & 0.775\pm0.031 & 0.792_{-0.035}^{+0.056}\\
		R_p\; [R_\earth] & 8.69\pm0.35 & 8.69\pm0.35 & 8.878_{-0.392}^{+0.628}\\
		\enddata
		\tablecomments{$^{\:\bm\dag}$ values estimated without using wavelet denoising and GP regression techniques (for comparison purpose).}
	\end{deluxetable*}
	
	\begin{deluxetable*}{LCCC}
		\tablecaption{Estimated physical parameters for WASP-95 b \label{tab:tab5}}
		\tablewidth{0pt}
		\tablehead{\colhead{Parameter} & \colhead{This work} & \colhead{Exp values$^{\:\bm\dag}$} & \colhead{\citet{2014MNRAS.440.1982H}}}
		\startdata
		\text{Transit parameters}\\
		T_0\; [BJD_{TDB}] & 2458326.505909\pm0.000027 & 2458326.505892\pm0.000048 & 2456338.45851\pm0.00024\\
		P\; [days] & 2.1846656\pm0.00000011 & 2.18466574\pm0.00000019 & 2.1846730\pm0.0000014\\
		b & 0.41984_{-0.00097}^{+0.00071} & 0.4197_{-0.0013}^{+0.0012} & 0.19_{-0.13}^{+0.21}\\
		R_\star/a & 0.168589_{-0.000045}^{+0.000039} & 0.168601_{-0.000073}^{+0.000083} & \nodata\\
		R_p/R_\star & 0.101692_{-0.000044}^{+0.000048} & 0.101694_{-0.000074}^{+0.000069} & 0.1025\pm0.0015\\
		\text{Limb-darkening coefficients}\\
		C_1 & 0.3345_{-0.0042}^{+0.0032} & 0.3332\pm0.0059 & \nodata\\
		C_2 & 0.2306_{-0.0040}^{+0.0037} & 0.2294_{-0.0079}^{+0.0067} & \nodata\\
		\text{Derived parameters}\\
		T_{14}\; [hr] & 2.8877_{-0.0015}^{+0.0017} & 2.8877_{-0.0015}^{+0.0012} & 2.784\pm0.024\\
		a/R_\star & 5.9316_{-0.0014}^{+0.0016} & 5.9316_{-0.0014}^{+0.0016} & \nodata\\
		i\; [deg] & 85.9411_{-0.0066}^{+0.0065} & 85.9411_{-0.0066}^{+0.0095} & 88.4_{-2.1}^{+1.2}\\
		M_p\; [M_J] & 1.206_{-0.067}^{+0.065} & 1.206_{-0.067}^{+0.065} & 1.13_{-0.04}^{+0.1}\\
		M_p\; [M_\earth] & 383.1_{-21.2}^{+20.7} & 383.1_{-21.2}^{+20.7} & 359.13_{-12.71}^{+31.78}\\
		T_{eq}\; [K] & 1692.6\pm40.4 & 1692.6\pm40.4 & 1570\pm50\\
		a\; [AU] & 0.0312\pm0.0022 & 0.0312\pm0.0022 & 0.03416\pm0.00083\\
		R_p\; [R_J] & 1.098\pm0.088 & 1.098\pm0.088 & 1.21\pm0.06\\
		R_p\; [R_\earth] & 12.31\pm0.99  & 12.31\pm0.99& 13.56\pm0.67\\
		\enddata
		\tablecomments{$^{\:\bm\dag}$ values estimated without using wavelet denoising and GP regression techniques (for comparison purpose).}
	\end{deluxetable*}
	
	\begin{deluxetable*}{LCCC}
		\tablecaption{Estimated physical parameters for WASP-156 b \label{tab:tab6}}
		\tablewidth{0pt}
		\tablehead{\colhead{Parameter} & \colhead{This work} & \colhead{Exp values$^{\:\bm\dag}$} & \colhead{\citet{2018AA...610A..63D}}}
		\startdata
		\text{Transit parameters}\\
		T_0\; [BJD_{TDB}] & 2458414.136153\pm0.000065 & 2458414.13622\pm0.00012 & 2454677.707\pm0.002\\
		P\; [days] & 3.8361603\pm0.00000048 & 3.83615987\pm0.00000089 & 3.836169\pm0.000003\\
		b & 0.2442_{-0.0073}^{+0.0061} & 0.240_{-0.010}^{+0.016} & \nodata\\
		R_\star/a & 0.07844_{-0.00015}^{+0.00016} 0.07843_{-0.00026}^{+0.00024} & 0.0781_{-0.0043}^{+0.0018}\\
		R_p/R_\star & 0.067654_{-0.000060}^{+0.000082} & 0.06766_{-0.00017}^{+0.00015} & 0.0685_{-0.0008}^{+0.0012}\\
		\text{Limb-darkening coefficients}\\
		C_1 & 0.4717_{-0.0043}^{+0.0040} & 0.4728_{-0.0061}^{+0.0077} & 0.461\pm0.002\\
		C_2 & 0.1344_{-0.0106}^{+0.0070} & 0.135_{-0.016}^{+0.015} & 0.0995_{-0.0074}^{+0.0066}\\
		\text{Derived parameters}\\
		T_{14}\; [hr] & 2.3926_{-0.0055}^{+0.0049} & 2.3930_{-0.0098}^{+0.0089} & 2.41_{-0.03}^{+0.04}\\
		a/R_\star & 12.748_{-0.027}^{+0.025} & 12.750_{-0.039}^{+0.043} & 12.8_{-0.7}^{+0.3}\\
		i\; [deg] & 88.902_{-0.028}^{+0.033} & 88.921_{-0.072}^{+0.049} & 89.1_{-0.9}^{+0.6}\\
		M_p\; [M_J] & 0.1303_{-0.0086}^{+0.0088} & 0.1303_{-0.0085}^+0.0088 & 0.128_{-0.009}^{+0.010}\\
		M_p\; [M_\earth] & 41.4_{-2.7}^{+2.8} & 41.4_{-2.7}^{+2.8} & 40.7_{-2.9}^{+3.2}\\
		T_{eq}\; [K] & 972.4\pm17.3 & 972.4\pm17.4 & 970_{-20}^{+30}\\
		a\; [AU] & 0.0451\pm0.0018 & 0.0451\pm0.0018 & 0.0453\pm0.0009\\
		R_p\; [R_J] & 0.554\pm0.034 & 0.554\pm0.034 & 0.51\pm0.02\\
		R_p\; [R_\earth] & 6.22\pm0.38 & 6.22\pm0.38 & 5.7\pm0.2
		\enddata
		\tablecomments{$^{\:\bm\dag}$ values estimated without using wavelet denoising and GP regression techniques (for comparison purpose).}
	\end{deluxetable*}
	
	\begin{deluxetable*}{lCC}
		\tablecaption{Best-fit GP regression model parameters}
		\label{tab:tab7}
		\tablewidth{0pt}
		\tablehead{\colhead{Target} & \colhead{$\mathrm{\alpha}$} & \colhead{$\mathrm{\tau}$}}
		\startdata
		KELT-7 b & 0.0002496_{-0.0000015}^{+0.0000014} & 0.002805_{-0.000016}^{+0.000012}\\
		HAT-P-14 b & 0.0003001_{-0.0000014}^{+0.0000016} & 0.003201_{-0.000015}^{+0.000020}\\
		WASP-29 b & 0.0004297_{-0.0000017}^{+0.0000015} & 0.002700_{-0.000013}^{+0.000016}\\
		WASP-95 b & 0.0003393_{-0.0000020}^{+0.0000021} & 0.002998_{-0.000050}^{+0.000056}\\
		WASP-156 b & 0.0005149_{-0.0000166}^{+0.0000127} & 0.002806_{-0.000086}^{+0.000077}
		\enddata
	\end{deluxetable*}
	
	\section{Target selection and observational data}\label{sec:obs}
	
	The TESS survey is aimed at detection of exoplanets around bright stars (9-15 mag), with the upper end limited only to exoplanets with significantly high transit depth. This is because the effective S/N decreases with the increase in magnitude. However, the studies involving precise estimation of physical properties require photometric observations with high S/N ($\gtrsim$ 100) and thereby limits the optimal magnitude for such studies ($\lesssim$ 12 mag).
	
	For our current study, we have selected five exoplanets, e.g., KELT-7 b, HAT-P-14 b, WASP-29 b, WASP-95 b and WASP-156 b, all of which are orbiting around bright stars with magnitude $<$ 12, and are within TESS survey field. We have found large uncertainties in their known parameter values with no significant follow-up studies. The properties of the target host stars used in our analysis are listed in Table \ref{tab:tab1}. In this table $K_{RV}$ is the radial velocity of the host star, $R_\star$ and $M_\star$ are its radius and mass respectively in term of solar radius $R_\sun$ and solar mass $M_\sun$, and $T_{eff}$ is its effective temperature. KELT-7 was observed in TESS sector 19, HAT-P-14 in 25 and 26, WASP-29 in 2 and 29, WASP-95 in 1 and 28, and WASP-156 in 4 and 31. We have obtained the TESS PDCSAP light-curves \citep{2014PASP..126..100S, 101086667697, 2012PASP..124..985S, 2017ksci.rept....2J} of these target stars from the public Mikulski Archive for Space Telescopes
	 (MAST)\footnote{https://mast.stsci.edu/portal/Mashup/Clients/Mast/Portal.html}. From these light-curves, we have identified 9, 11, 11, 20 and 12 full transit observations for KELT-7 b, HAT-P-14 b, WASP-29 b, WASP-95 b and WASP-156 b respectively.
	
	\section{Data analysis and modeling}\label{sec:analysis}
	
	In this section, we describe the two most important aspects of this study, i.e. the noise treatment and the modeling of the transit light-curves. For this part of the study, we have used our semi-automated software package, written in Python and used in our previous investigations \citep{2019AJ....158...39C, Saha_2021}.
	
	At first, we have sliced the long continuous TESS light-curves into smaller segments concentrated around the observed full-transit events to obtain the transit light-curves. We have used only full-transit observations in this study to account for the best possible accuracy. To eliminate any long-term variability effects in these light-curves, we have treated them with the baseline correction method following the same procedure as in \citet{Saha_2021}.
	
	To reduce the small-scale time-uncorrelated fluctuations in the light-curves due to various instrumental effects, we have used the wavelet denoising technique \citep{Donoho1994IdealDI, 806084, WaveletDenoise2012, 2019AJ....158...39C, Saha_2021}. For this purpose, we have used the PyWavelets \citep{Lee2019} python package for the wavelet based operations and followed the procedure described in \citet{Saha_2021}. We have shown the transit light-curves before and after wavelet denoising in Figures \ref{fig:fig0}-\ref{fig:fig5}.
	
	We have reduced the correlated noise components in the light-curves due to stellar activities and pulsations by modeling them simultaneously along with the transit signals using the GP regression method \citep{2006gpml.book.....R, 2015ApJ...810L..23J, 2019MNRAS.489.5764P, 2020AA...634A..75B, 2019AJ....158...39C, Saha_2021}. Following \citet{Saha_2021}, we have used the formalism described in \citet{2006gpml.book.....R} for noisy observations. We have shown the best-fit GP regression models over-plotted on the correlated noise components and the reduced mean residuals following the GP regression in Figures \ref{fig:fig0}-\ref{fig:fig5}.
	
	To model the transit signals, we have followed the analytical transit model formulation given by \citet{2002ApJ...580L.171M}. This formulation incorporates the limb-darkening effects of the host stars by using a quadratic limb-darkening formula. In order to estimate the transit parameters for our target exoplanets, we have used the MCMC sampling technique by incorporating the Metropolis-Hastings algorithm \citep{1970Bimka..57...97H} for the simultaneous modeling of the transit light-curves. A detailed description of the this technique can be found in \citet{Saha_2021}. We have shown the best-fit transit models over-plotted on the observed and wavelet denoised light-curves in the Figures \ref{fig:fig0}-\ref{fig:fig5}. To understand the improvements in the values of various physical parameters estimated by using our critical noise reduction techniques, i.e. wavelet denoising and GP regression, we have also modeled the transit light-curves of our target exoplanets directly without using these techniques.
	
	One of the estimated parameters from the modeling of our transit light-curves is the mid-transit times $T_C$. These values of $T_C$ are used to model the linear ephemeris of these exoplanets with two free parameters, $T_0$ and $P$, using the MCMC sampling technique. Following the relations given in \citet{Saha_2021}, we have derived the values of the remaining physical parameters from the estimated parameter values and the adopted host star properties as listed in Table \ref{tab:tab1}. We have tabulated all the derived physical parameters of our target exoplanets in Tables \ref{tab:tab2}-\ref{tab:tab6}. The explanation of all the terms are given in the Table \hyperref[tab:tab9]{A1} in the Appendix. We have also listed the best-fit GP regression model parameters in Table \ref{tab:tab7}, while the estimated mid-transit times with O-C deviations from the best-fit linear ephemeris are provided in Table \ref{tab:tab8}.
	
	\section{Discussion}\label{sec:dis}
	
	In this study, we have conducted transit photometric follow-up analysis of four hot Jupiters, KELT-7 b, HAT-P-14 b, WASP-29 b, WASP-95 b, and a hot Neptune, WASP-156 b, using the photometric observations from the TESS space-based telescope, and using the critical noise reduction techniques, i.e. wavelet denoising and GP regression. The main purpose have been to estimate the various physical properties of these exoplanets more precisely and accurately.
		
	Starting with the photometric observations from TESS, having a constant cadence of 120 seconds of exposure time means a gradual decrease in S/N with the increase in magnitude \citep{2010ApJ...713L..97B}. The visible difference in S/N is well noticeable in the KELT-7, which is the brightest among all our target stars and the WASP-156, the faintest one. However, in the present work, the absence of any fluctuations in the light-curves from the atmospheric effects has given us an advantage over the previous studies of our target exoplanets, all of which were based on observations using ground-based facilities.
	
	The small scale fluctuations in the light-curves caused by various instrumental effects that are uncorrelated in time, are especially noticeable because of the absence atmospheric effects. We have reduced them using the same wavelet denoising technique which has been found to be efficient in our previous studies of transiting exoplanets observed by using ground-based facilities. The time-correlated noise components due to stellar activities and pulsations were also reduced by applying GP regression method. It is worth mentioning here that the amplitude of these correlated noise components in TESS observations is small compared to those from ground-based observations. This is because of the fact that unlike their ground-based counterparts, the space-based observations do not require differential photometric method in order to tackle the effects of air-mass and atmospheric variability. This is also the reason why we have used a single set of GP regression parameters for each of our targets.
	
	We have estimated the photometric precision of the lightcurves at each step of our critical noise treatment algorithm, by computing the Combined Differential Photometric Precision (CDPP) noise metric \citep{2011ApJS..197....6G, 2016PASP..128g5002V} using the \textit{estimate\_cdpp} module of \textit{Lightkurve} \citep{2018ascl.soft12013L} package, and listed in the Table \ref{tab:tab10}. We can see a small improvement in the photometric precision due to the wavelet denoising technique and a significant improvement due to the GP regression technique. The small improvement due to wavelet denoising is desirable, as a large correction at this step will lead to a over-smoothing of the lightcurves, resulting in deformation of the transit signals. On the other hand, the GP regression is performed simultaneously alongwith modeling for the transit signals, and thus a large correction has no impact in deforming the lightcurves.
	
	To appreciate the improvements brought upon by the critical noise reduction techniques, we have also modeled the raw light-curves of our targets without using any of those techniques. We have then compared the estimated values of various physical parameters with those derived after using the critical noise reduction techniques (see Table \ref{tab:tab2}). We can clearly notice the improvements in the uncertainty limits in the estimated and directly derived transit parameters achieved by using the critical noise reduction techniques. This proves that the noise components can severely affect the precision and overall accuracy of the estimated parameters when left untreated, and therefore require critical noise reduction methods to reduce them efficiently. However, there is no difference in the uncertainty limits of the estimated values for the parameters that require the adopted properties of the host stars. This is because of the large uncertainties in those stellar parameters.
	
	To understand the improvements in the values of the physical parameters of the target exoplanets derived by us in the present work, we have also compared them with their estimated values from previous studies. We have found a few order of magnitude improvements in the values of the estimated transit parameters and parameters derived from them directly. This is expected and can be attributed to a combination of very high S/N photometric observations from TESS and the state-of-the-art critical noise treatment algorithm used in our study. Only the uncertainty limits in the estimated values for the orbital periods of KELT-7 b and HAT-P-14 b are comparatively large. This is because of the fact that the follow-up observations of these two planets are over a very limited number of transit epochs (9 and 11 respectively). This can be improved by further follow-up studies and combining the results with the precise values for mid-transit times estimated in this study (see Table \ref{tab:tab8}). On the other hand, the estimated values of the parameters derived from the stellar parameters, do not have significant improvements over uncertainty limits because of very high uncertainties in the values of the adopted stellar properties.
	
	Overall, our work has resulted into a more precise and accurate estimation of the physical parameters for our target exoplanets as compared to that reported in previous studies. This improvements are attributed to a better quality of photometric data from TESS and reduction of various noise components using a critical noise treatment algorithm.
	
	\startlongtable
	\begin{deluxetable*}{lCCC}
		\tablecaption{Estimated mid-transit times}
		\label{tab:tab8}
		\tablewidth{0pt}
		\tablehead{\colhead{Target} & \colhead{Epoch} & \colhead{$\mathrm{T_C\;[BJD_{TDB}]}$} & \colhead{$\mathrm{O-C\;[sec]}$}}
		\startdata
		KELT-7 b & 0 & 2458816.517884_{-0.000102}^{+0.000129} & -43.23\\
		& 1 & 2458819.253384_{-0.000068}^{+0.000083} & 18.61\\
		& 2 & 2458821.987897_{-0.000073}^{+0.000085} & -4.78\\
		& 3 & 2458824.723077_{-0.000083}^{+0.000059} & 29.39\\
		& 4 & 2458827.457495_{-0.000081}^{+0.000087} & -2.24\\
		& 5 & 2458830.192210_{-0.000071}^{+0.000059} & -8.20\\
		& 6 & 2458832.926792_{-0.000073}^{+0.000081} & -25.68\\
		& 7 & 2458835.661988_{-0.000080}^{+0.000091} & 9.94\\
		& 8 & 2458838.396687_{-0.000057}^{+0.000071} & 2.57\\
		\hline
		HAT-P-14 b & 0 & 2458984.653102_{-0.000093}^{+0.000094} & -26.92\\
		& 1 & 2458989.281403_{-0.000071}^{+0.000078} & 26.23\\
		& 2 & 2458993.908200_{-0.000086}^{+0.000072} & -50.61\\
		& 3 & 2458998.535608_{-0.000090}^{+0.000078} & -74.64\\
		& 4 & 2459003.165288_{-0.000081}^{+0.000085} & 97.57\\
		& 5 & 2459007.792394_{-0.000065}^{+0.000056} & 47.44\\
		& 6 & 2459012.419598_{-0.000067}^{+0.000082} & 5.80\\
		& 7 & 2459017.046880_{-0.000064}^{+0.000075} & -29.13\\
		& 8 & 2459021.675296_{-0.000061}^{+0.000080} & 33.94\\
		& 9 & 2459026.301697_{-0.000061}^{+0.000076} & -77.07\\
		& 10 & 2459030.930598_{-0.000083}^{+0.000073} & 27.82\\
		\hline
		WASP-29 b & 0 & 2458356.415209_{-0.000057}^{+0.000069} & 29.26\\
		& 1 & 2458360.337309_{-0.000057}^{+0.000058} & -23.58\\
		& 2 & 2458364.260001_{-0.000081}^{+0.000073} & -25.37\\
		& 4 & 2458372.105822_{-0.000107}^{+0.000073} & 8.96\\
		& 5 & 2458376.028072_{-0.000075}^{+0.000092} & -31.00\\
		& 6 & 2458379.951496_{-0.000071}^{+0.000056} & 30.50\\
		& 187 & 2459089.962194_{-0.000092}^{+0.000065} & 12.68\\
		& 188 & 2459093.884380_{-0.000074}^{+0.000077} & -32.74\\
		& 189 & 2459097.807400_{-0.000071}^{+0.000064} & 6.21\\
		& 191 & 2459105.653298_{-0.000080}^{+0.000079} & 34.71\\
		& 192 & 2459109.575601_{-0.000083}^{+0.000091} & -0.57\\
		\hline
		WASP-95 b & 0 & 2458326.506113_{-0.000069}^{+0.000085} & 17.64\\
		& 1 & 2458328.690901_{-0.000086}^{+0.000082} & 28.26\\
		& 2 & 2458330.874826_{-0.000100}^{+0.000075} & -35.69\\
		& 3 & 2458333.060402_{-0.000091}^{+0.000080} & 42.88\\
		& 5 & 2458337.428737_{-0.000082}^{+0.000078} & -43.18\\
		& 7 & 2458341.798552_{-0.000085}^{+0.000122} & -1.37\\
		& 8 & 2458343.983001_{-0.000105}^{+0.000110} & -20.09\\
		& 9 & 2458346.167908_{-0.000082}^{+0.000067} & 70.29\\
		& 11 & 2458350.537197_{-0.000059}^{+0.000065} & -2.91\\
		& 12 & 2458352.721972_{-0.000069}^{+0.000081} & 6.53\\
		& 337 & 2459062.738413_{-0.000060}^{+0.000064} & 15.67\\
		& 338 & 2459064.922486_{-0.000069}^{+0.000069} & -35.58\\
		& 339 & 2459067.107641_{-0.000114}^{+0.000081} & 6.70\\
		& 340 & 2459069.292621_{-0.000063}^{+0.000066} & 33.86\\
		& 341 & 2459071.476797_{-0.000065}^{+0.000070} & -8.41\\
		& 343 & 2459075.846513_{-0.000076}^{+0.000081} & 24.81\\
		& 344 & 2459078.031202_{-0.000087}^{+0.000069} & 26.88\\
		& 345 & 2459080.215201_{-0.000062}^{+0.000069} & -30.73\\
		& 346 & 2459082.400225_{-0.000094}^{+0.000071} & 25.42\\
		& 347 & 2459084.584804_{-0.000064}^{+0.000076} & -7.3\\
		\hline
		WASP-156 b & 0 & 2458414.135676_{-0.000127}^{+0.000134} & -41.22\\
		& 1 & 2458417.971972_{-0.000197}^{+0.000172} & -29.49\\
		& 2 & 2458421.809599_{-0.000179}^{+0.000144} & 97.25\\
		& 3 & 2458425.644830_{-0.000155}^{+0.000139} & 16.94\\
		& 4 & 2458429.480008_{-0.000138}^{+0.000160} & -67.93\\
		& 5 & 2458433.317222_{-0.000128}^{+0.000154} & 23.10\\
		& 191 & 2459146.842429_{-0.000127}^{+0.000122} & -29.18\\
		& 192 & 2459150.678210_{-0.000193}^{+0.000180} & -61.88\\
		& 193 & 2459154.514789_{-0.000169}^{+0.000164} & -25.70\\
		& 195 & 2459162.188300_{-0.000172}^{+0.000153} & 77.08\\
		& 196 & 2459166.022691_{-0.000146}^{+0.000176} & -75.71\\
		& 197 & 2459169.860999_{-0.000156}^{+0.000152} & 109.79\\
		\enddata
	\end{deluxetable*}
	
	\begin{deluxetable*}{lCCC}
		\tablecaption{Estimated CDPP for lightcurves}
		\label{tab:tab10}
		\tablewidth{0pt}
		\tablehead{\colhead{Target} & & \colhead{CDPP} \\
			& \colhead{PDCSAP lightcurves} & \colhead{After wavelet denoising} & \colhead{After GP regression}}
		\startdata
		KELT-7 b & 138.0 & 134.5 & 40.3\\
		HAT-P-14 b & 227.0 & 222.7 & 117.2\\
		WASP-29 b & 243.4 & 231.1 & 131.1\\
		WASP-95 b & 245.4 & 241.7 & 97.8\\
		WASP-156 b & 328.2 & 319.5 & 189.7
		\enddata
	\end{deluxetable*}
	
	\section{Conclusion}\label{sec:con}
	
	In this work, we have critically analyzed TESS transit photometric data of five already discovered exoplanets, e.g., KELT-7 b, HAT-P-14 b, WASP-29 b, WASP-95 b and WASP-156 b. Being a space based telescope, the TESS observations are unaffected by Earth's atmospheric interferences. Also, since all the target host stars in our study were brighter than 12 magnitude, we have found high S/N in all the transit light-curves.
	
	In order to improve the quality of the transit signal in the light-curves, we have reduced the noise components due to various sources by using a state-of-the-art critical noise treatment algorithm that includes wavelet denoising and GP regression methods. Thus, we have demonstrated the efficiency of these techniques in improving the precision and overall accuracy of the estimated parameters.
	
	The modeling of the transit light-curves has used the MCMC sampling technique, which has provided extremely precise estimation of the physical parameters for our target exoplanets. This can be attributed to both the quality of the photometric observations from TESS and the critical noise treatment algorithm used in this study. We have compared our estimated parameter values with those from the previous studies to show the improvements in precision and overall accuracy. The work establishes the power of space-based photometric follow-up studies when combined with a critical noise treatment approach. 
	
	We thank Aritra Chakrabarty for providing helpful inputs on the numerical aspects of this work. We are thankful to the anonymous reviewer for a critical reading of the manuscript and for providing many useful comments and suggestions. Some of the computational results reported in this work were performed on the high performance computing facility (NOVA) of IIA, Bangalore. We are thankful to the computer division of Indian Institute of Astrophysics for the help and co-operation extended to us.
	
	This paper includes data collected by the TESS mission, which are publicly available from the Mikulski Archive for Space Telescopes (MAST). We acknowledge the use of public TOI Release data from pipelines at the TESS Science Office and at the TESS Science Processing Operations Center. Funding for the TESS mission is provided by NASA’s Science Mission directorate. Support for MAST is provided by the NASA Office of Space Science via grant NNX13AC07G and by other grants and contracts. This research made use of Lightkurve, a Python package for Kepler and TESS data analysis.
	
	\bibliography{ms}{}
	\bibliographystyle{aasjournal}
	
	\appendix
	
	\begin{deluxetable*}{Lc}[h]
		\tablecaption{Meaning of all the parametric terms presented in Tables \ref{tab:tab2}-\ref{tab:tab8}}
		\label{tab:tab9}
		\tablenum{A1}
		\tablewidth{0pt}
		\tablehead{\colhead{Term} & \colhead{Explanation}}
		\startdata
		T_0\; [BJD_{TDB}] & Initial transit time in BJD-TDB\\
		P\; [days] & Orbital period in days\\
		R_\star/a & Scaled stellar radius\\
		R_p/R_\star & Ratio of planet to stellar radius\\
		C_1 & Limb darkening coefficient (linear term)\\
		C_2 & Limb darkening coefficient (quadratic term)\\
		T_{14}\; [hr] & Transit duration in hours\\
		a/R_\star & Scale parameter\\
		i\; [deg] & Orbital inclination angle in degree\\
		M_p\; [M_J] & Mass of the planet in Jupiter-mass\\
		M_p\; [M_\earth] & Mass of the planet in Earth-mass\\
		T_{eq}\; [K] & Equilibrium temperature of the planet in Kelvin\\
		a\; [AU] & Semi-major axis in AU\\
		R_p\; [R_J] & Radius of planet in Jupiter-radius\\
		R_p\; [R_\earth] & Radius of planet in Earth-radius\\
		\alpha & GP regression parameter (signal standard deviation)\\
		\tau & GP regression parameter (characteristic length scale)\\
		T_C\;[BJD_{TDB}] & Mid-transit time in BJD-TDB\\
		O-C\;[sec] & Deviation of mid-transit time for best-fit linear ephemeris in seconds
		\enddata
	\end{deluxetable*}

	\begin{figure*}[h]
		\centering
		\includegraphics[width=\linewidth]{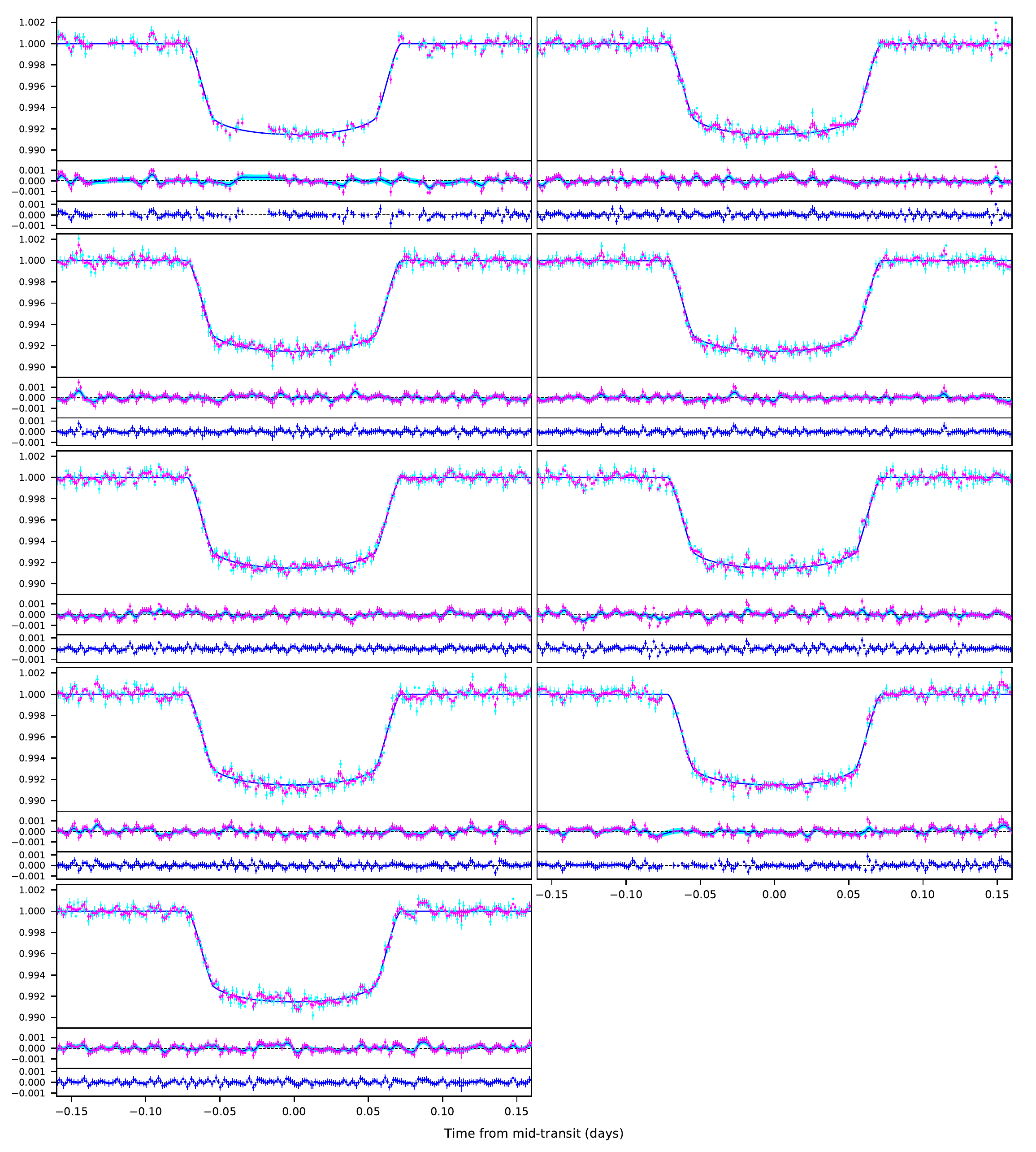}
		\caption{All observed and best-fit model light-curves for KELT-7 b, with descriptions same as Figure \ref{fig:fig0}}
		\label{fig:fig1}
	\end{figure*}
	
	\begin{figure*}[h]
		\centering
		\includegraphics[width=\linewidth]{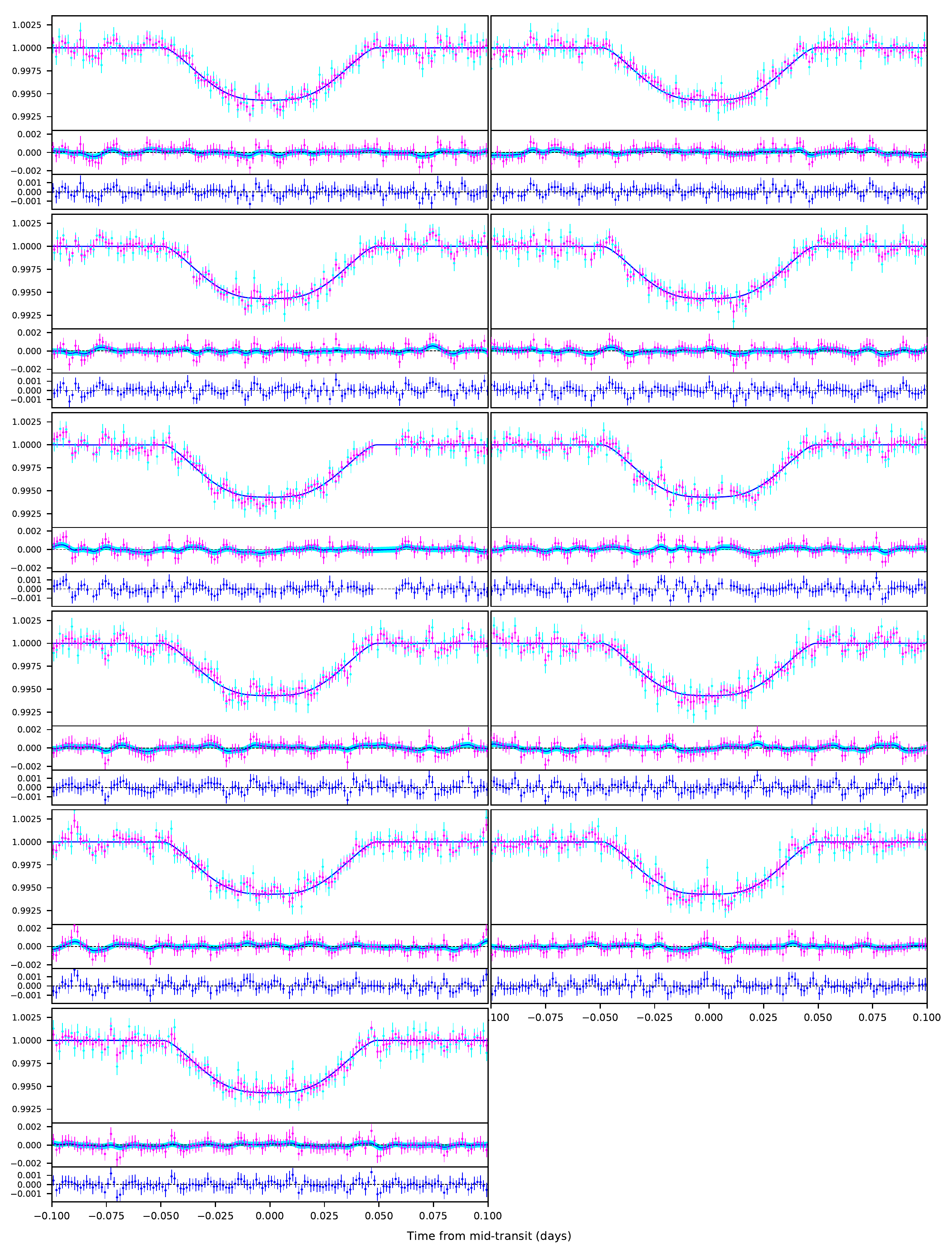}
		\caption{Same as Figure \ref{fig:fig1}, but for HAT-P-14 b}
		\label{fig:fig2}
	\end{figure*}
	
	\begin{figure*}[h]
		\centering
		\includegraphics[width=\linewidth]{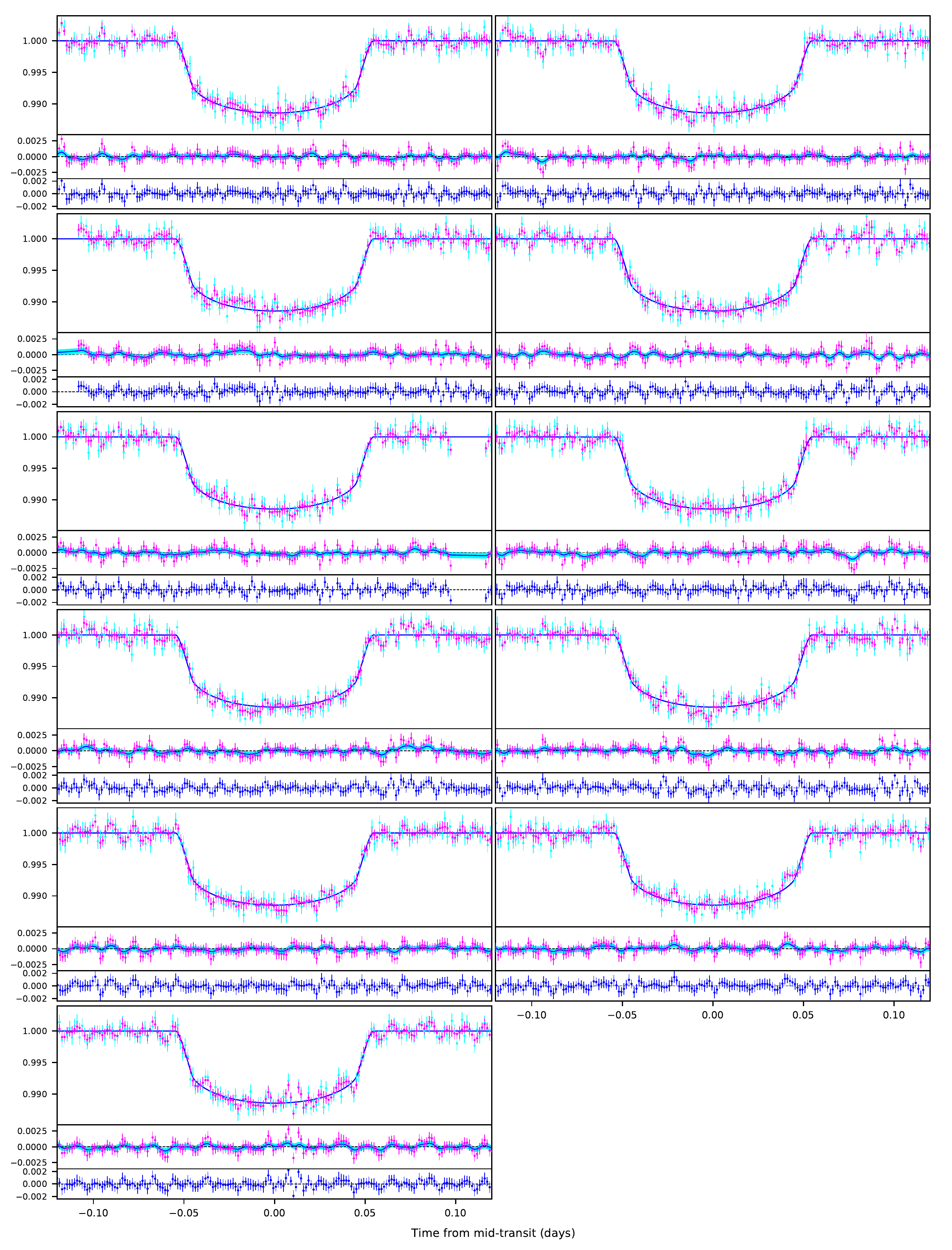}
		\caption{Same as Figure \ref{fig:fig1}, but for WASP-29 b}
		\label{fig:fig3}
	\end{figure*}
	
	\begin{figure*}[h]
		\centering
		\includegraphics[width=\linewidth]{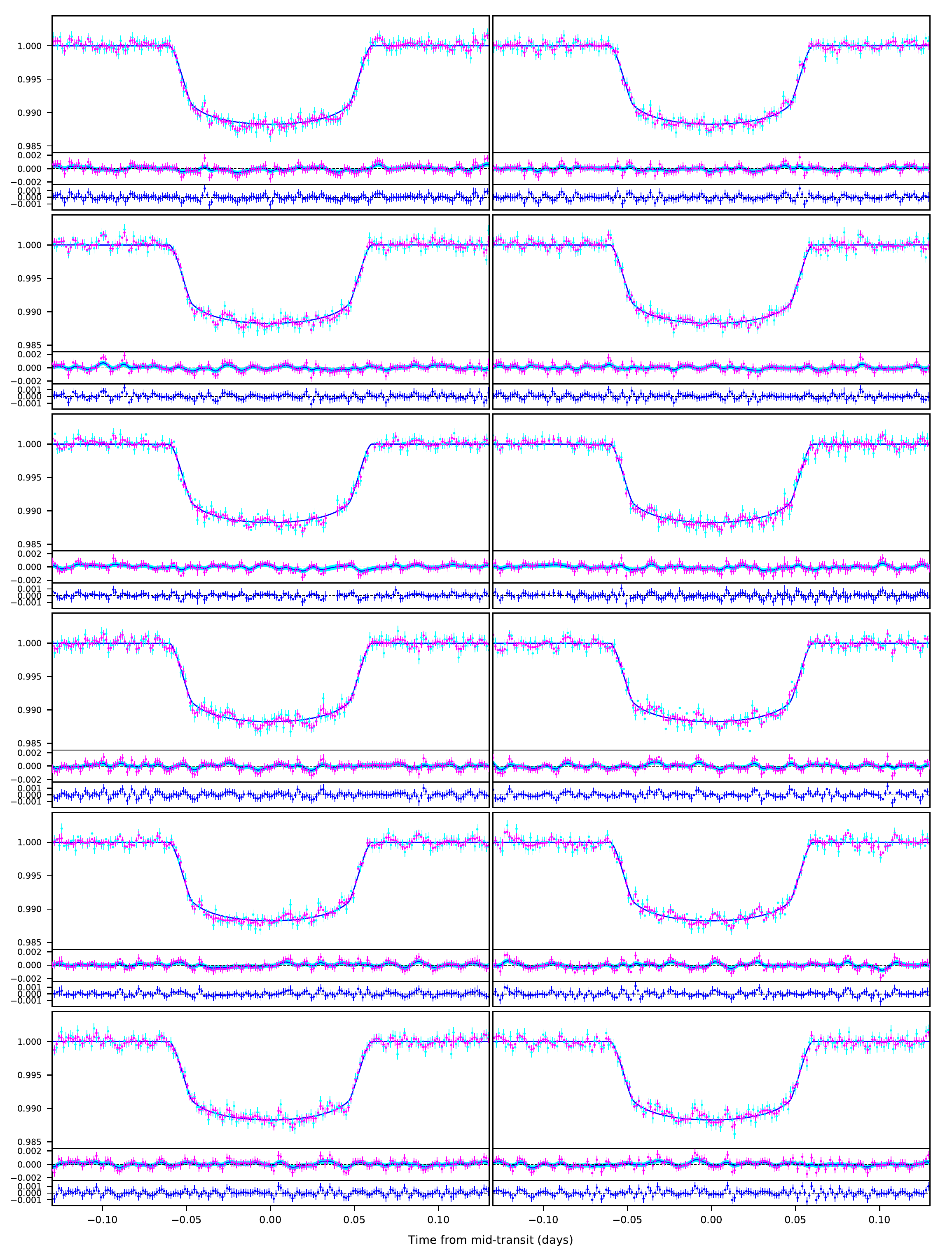}
		\caption{Same as Figure \ref{fig:fig1}, but for WASP-95 b}
		\label{fig:fig41}
	\end{figure*}
	
	\begin{figure*}[h]
		\centering
		\figurenum{5}
		\includegraphics[width=\linewidth]{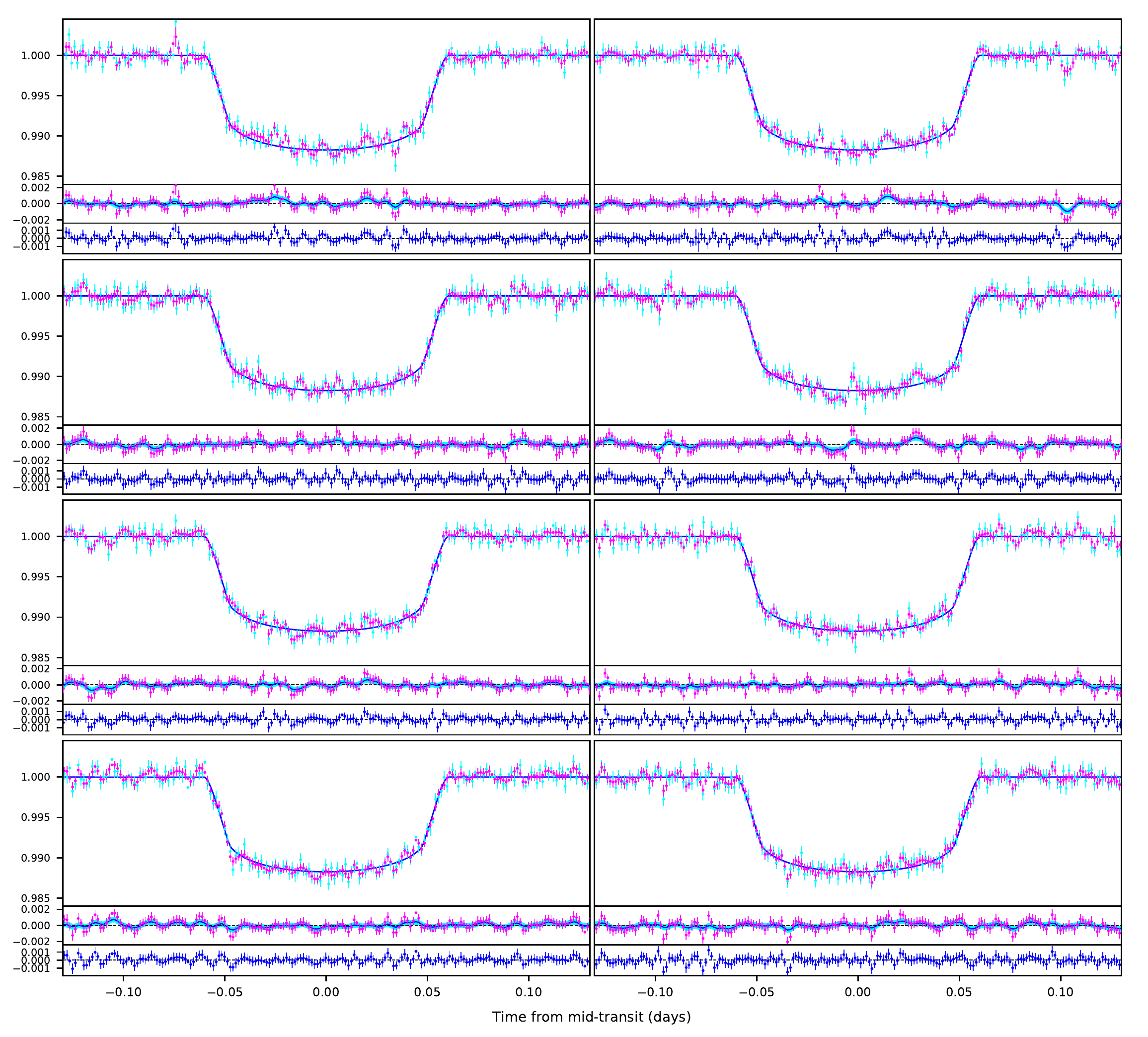}
		\caption{(Continued)}
		\label{fig:fig42}
	\end{figure*}
	
	\begin{figure*}[h]
		\centering
		\includegraphics[width=\linewidth]{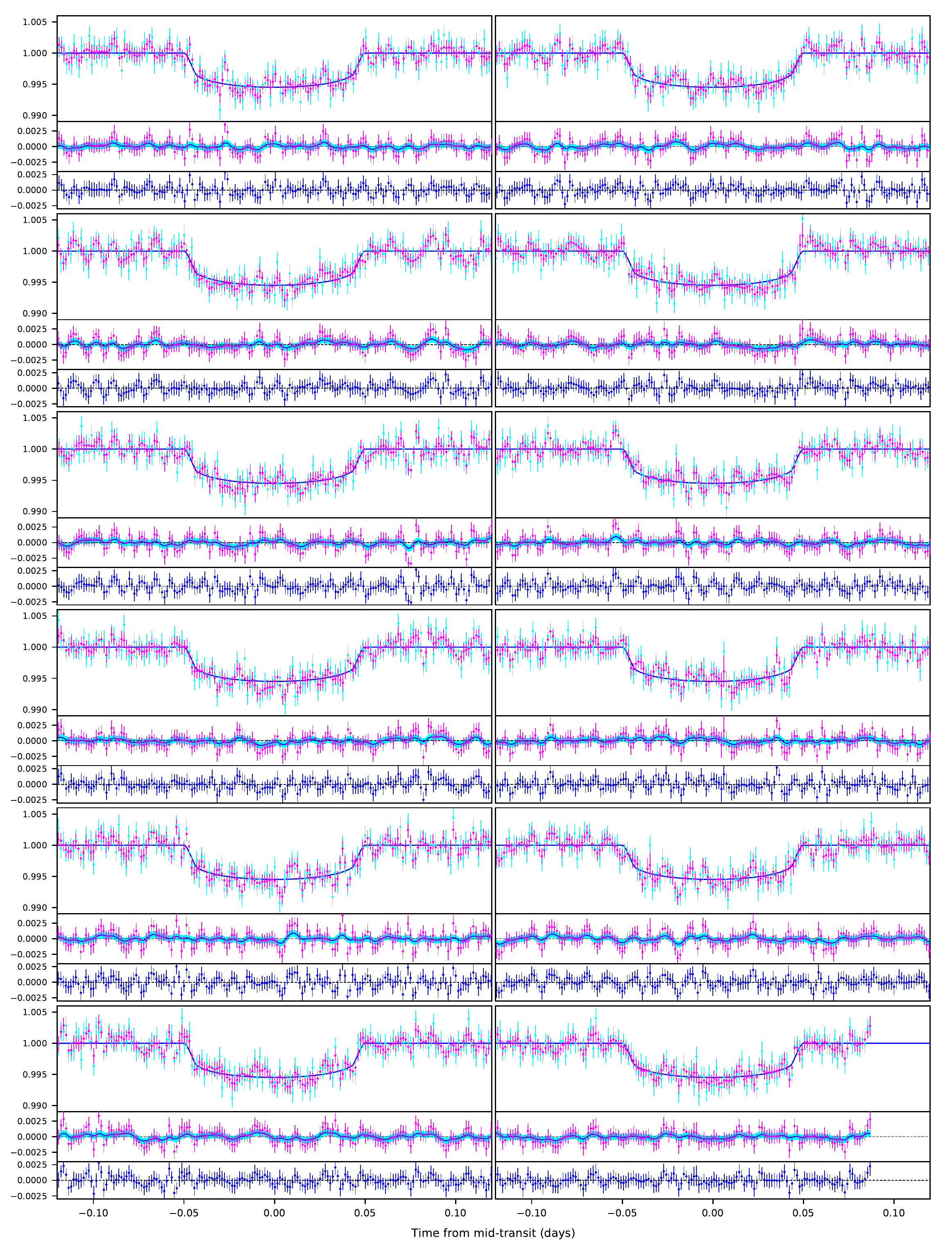}
		\caption{Same as Figure \ref{fig:fig1}, but for WASP-156 b}
		\label{fig:fig5}
	\end{figure*}
	
\end{document}